\documentclass[conf]{IEEEtran}
\ifCLASSINFOpdf
\else
\fi
\usepackage{epsfig,graphics,graphicx,subfigure,amssymb,amstext,amsmath,algorithm,algorithmic,wrapfig,multirow}
\usepackage{array}
\usepackage{cite}
\usepackage{url}
\usepackage{times}
\usepackage{lipsum}
\usepackage{float}
\usepackage{placeins}
\usepackage{textcomp}
\usepackage[font=small,bf]{caption}
\usepackage{hyperref}
\usepackage{flushend}
\usepackage{xcolor}
\usepackage{tabularx}
\usepackage{bm}
\usepackage{enumerate}
\usepackage{tikz}
\usepackage{pgfplots}
\usepackage{enumerate}
\usepackage{epstopdf}
\usetikzlibrary{shapes,arrows}
\usepackage[utf8]{inputenc}
\usepackage{mathtools}

\def\beq{\begin{equation}}
\def \eeq{\end{equation}}
\def\beqa{\begin{eqnarray}}
\def\eeqa{\end{eqnarray}}
\def\beqan{\begin{eqnarray*}}
\def\eeqan{\end{eqnarray*}}

\def\C{{\mathbb{C}}}

\def\arr{\rightarrow}
\def\Exp{\mathbb{E}}

\def\tm1{t\! - \! 1}
\def\tp1{t\! + \! 1}

\def\nbf{\mathbf{n}}

\def\vbf{\mathbf{v}}

\def\wbf{\mathbf{w}}

\def\ybf{\mathbf{y}}
\def\zbf{\mathbf{z}}

\def\Hbf{\mathbf{H}}

\def\Ibf{\mathbf{I}}

\def\Qbf{\mathbf{Q}}

\newif\ifconf
\conffalse

\newif\ifthreebitq
\threebitqtrue

\newif\ifonecol
\onecolfalse

\IEEEoverridecommandlockouts

\renewcommand{\footnoterule}{%
  \kern -3pt
  \hrule width \columnwidth height 0.5pt
  \kern 3pt
}

\begin{document}
\title{5{G} Millimeter Wave Cellular System Capacity 
with Fully Digital Beamforming}

\author{Sourjya~Dutta,
        C. Nicolas~Barati,
        Aditya Dhananjay,
        and~Sundeep~Rangan \\
        NYU Wireless, Tandon School of Engineering, New York University, Brooklyn, NY.
\thanks{The authors are with NYU Wireless, NYU Tandon School of Engineering, Brooklyn,
NY, 11201, USA. e-mail: \{sdutta, cbn228, srangan\}@nyu.edu, aditya@courant.nyu.edu.
}
\thanks{
Copyright 2017 SS\&C. Published in the Proceedings of 51st Asilomar Conference on Signals, Systems, and Computers, Oct. 2017, Pacific Grove, CA, USA.}

\thanks{This work was supported in part by NSF grants 1302336, 1564142, and 1547332, NSA, NIST, and the affiliates members of NYU WIRELESS.
}
}


\maketitle

\begin{abstract}
Due to heavy reliance of millimeter-wave (mmWave) wireless systems
on directional links,
Beamforming (BF) with high-dimensional arrays is essential 
for cellular systems in these frequencies.
How to perform the array processing in a power efficient manner is 
a fundamental challenge.
Analog and hybrid BF require fewer analog-to-digital converters (ADCs),
but can only communicate in a small number of directions at a time,
limiting directional search, spatial multiplexing and control signaling.
Digital BF enables flexible spatial processing, but must be operated 
at a low quantization resolution to stay within reasonable power levels.
This paper presents a simple additive white Gaussian noise (AWGN) model
to assess the effect of low-resolution quantization of cellular system capacity.
Simulations with this model reveal that at moderate resolutions
(3-4 bits per ADC), there is negligible loss in downlink 
cellular capacity from quantization. In essence, the low-resolution
ADCs limit the high SNR, where cellular systems typically do not operate.
The findings suggest that low-resolution fully digital BF architectures
can be power efficient, offer greatly enhanced control plane functionality and comparable data plane performance to analog BF.


     
\end{abstract}


\IEEEpeerreviewmaketitle

\section{Introduction}
\IEEEPARstart{T}{he} 
need for more bandwidth, driven by ever higher demand, has brought millimeter wave (mmWaves) communication into the spotlight as a solid candidate technology for the 5\textsuperscript{th} generation (5G) wireless communications.
By offering large blocks of contiguous spectrum, mmWave presents a unique
opportunity to overcome the bandwidth crunch problem in lower frequency bands \cite{RanRapE:14}.

High isotropic path loss at mmWave frequencies necessitates  
the reliance on antenna arrays with large number of elements. These arrays overcome the path loss by high directional gains through beamforming (BF). Thus, a transmitter--receiver (Tx--Rx) pair uses a multiplicity of antennas to focus energy in a particular direction to
meet a target link budget.
A basic question is how to perform the high-dimensional array processing in a
power-efficient manner, particularly for handheld user equipments (UEs).

\begin{figure}
\centering
\begin{tikzpicture}[scale=0.35]
\draw[thick] (0,0) rectangle (2,10) node [align=center] at (1,5) {$\mathbf{w_{TX}}$};
\draw[thick] (18,0) rectangle (20,10) node [align=center] at (19,5) {$\mathbf{w_{RX}}$};

\foreach \y in {1,5,7,9} {
\draw[thick] (2,\y) -- (3,\y); 
\draw[thick] (3,\y-0.5) rectangle (4,\y+0.5) node [align=center] at (3.5, \y) {\tiny D/A};

\draw[thick] (18,\y) -- (17,\y); 
\draw[thick] (17,\y-0.5) rectangle (16,\y+0.5) node [align=center] at (16.5, \y) {\tiny A/D};

\draw[thick] (4,\y) -- (5,\y); \draw[thick] (16,\y) -- (15,\y);
\draw[thick] (5.5,\y) circle (0.5) node [align=center] at (5.5,\y) {\tiny $\times$}; 
\draw[thick] (14.5,\y) circle (0.5) node [align=center] at (14.5,\y) {\tiny $\times$}; 

\draw[thick] (6,\y) -- (7,\y); \draw[thick] (14,\y) -- (13,\y);
\draw[thick] (7,\y-0.5) rectangle (8,\y+0.5) node [align=center] at (7.5, \y) {\tiny PA};
\draw[thick] (13,\y-0.5) rectangle (12,\y+0.5) node [align=center] at (12.5, \y) {\tiny PA};
\draw[thick] (8,\y) -- (8.5,\y); \draw[thick] (12,\y) -- (11.5,\y);

\draw[thick] (8.5,\y) -- (8.6,\y+0.1) -- (8.6,\y-0.1) --cycle;
\draw[thick] (11.5,\y) -- (11.4,\y+0.1) -- (11.4,\y-0.1) --cycle;
}

\draw[->, thick] (-1,5) -- (0,5) node [align=center] at (-0.5,5.5) {$x$};
\draw[->, thick] (20,5) -- (21,5) node [align=center] at (20.5,5.5) {$\hat{y}$};

\node [align=center] at (6, 3) {$\vdots$}; \node [align=center] at (14, 3) {$\vdots$};

\node [align=center] at (10,5) {\large $H$}; \node [align=center] at (8.6,0.5) {\tiny$N_{\rm TX}$};
\node [align=center] at (8.6,8.5) {\tiny$1$}; \node [align=center] at (8.6,6.5) {\tiny$2$};
 \node [align=center] at (8.6,4.5) {\tiny$3$};

\node [align=center] at (11.4,0.5) {\tiny$N_{\rm RX}$};
\node [align=center] at (11.4,8.5) {\tiny$1$}; \node [align=center] at (11.4,6.5) {\tiny$2$}; \node [align=center] at (11.4,4.5) {\tiny$3$};

\end{tikzpicture}
\caption{Fully digital transmitter (left) and receiver (right).}
\label{fig:txrx}
\end{figure}
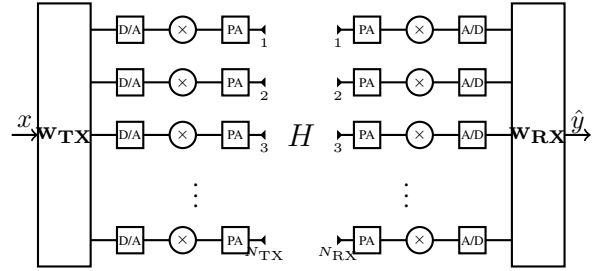

Most current commercial mmWave designs use
\emph{analog} or \emph{hybrid beamforming}.
In this case, beamforming is performed in RF (or IF) 
through a bank of phase shifters -- one per antenna element.
This architecture reduces the power consumption by using only a pair of 
analog to digital converters (ADC) and digital to analog converters (DAC) at the Rx and Tx respectively per digital stream. 
Although, the power consumption is reduced, the Tx and Rx can only transmit in one direction per digital stream at a given time \cite{KhanPi:11-CommMag}. 
In contrast, in \emph{fully digital architectures} \cite{Zhang:05},
as shown in Fig. \ref{fig:txrx}, beamforming is performed
in baseband.
Each RF chain has a pair of ADCs at the Rx and DACs at the Tx.
This enables the transceiver to direct beams at infinitely many directions at the
same time.
The fully digital architecture enables greatly enhanced
spatial flexibility.
However, to maintain similar power consumption levels as analog BF,
fully digital arrays must typically operate at low quantization levels.
Hence, there is a fundamental tradeoff between directional search and 
spatial multiplexing on the one hand and quantization noise on the other hand.

It is now well-known that the spatial flexibility afforded by
fully digital architectures is invaluable in the \emph{control plane}.
For example, it is shown in \cite{barati2015directional,barati2016initial} that
digital BF can reduce control plane latency by orders of magnitude.
Moreover, digital beamforming enables 
multi-stream communication leading to the effective use of the spatial degrees of freedom.
Additionally, with digital beamforming frequency division multiple access (FDMA) scheduling is possible enabling much more efficient transmission of short data and control packets
\cite{dutta2017frame}. 
 

The main contribution of this work is the analysis of low resolution fully-digital receiver architectures at mmWave frequencies in the \emph{data plane}. 
There has been significant information theoretic studies
\cite{Singh2009Limits,oner2015adc,mo2014channel,mo2014high}
on point-to-point capacity with low-resolution limits. The focus of this work is to study low-resolution effects in multi-user cellular environments. To this end, we first derive and validate a simple additive white quantization noise (AWQN) link-layer model for the received signal to noise and interference ratio (SINR) when the receiver employs low resolution ADCs.
Secondly, we apply the AWQN model along with 
detailed simulations to study the effect of low quantization on the SINR and rate of mmWave cellular users in a multi-user scenario. 

The simulations reveal that with the use of low resolution ADCs 
(e.g.\ 3--4 bits per ADC), 
\emph{there is negligible loss in the overall cellular system capacity}.
Moreover, at these quantization rates, fully digital BF architectures can be implemented
with \emph{lower power} than current analog solutions.
Hence, fully digital BF architectures at these resolutions may offer
greatly enhanced control plane functionality, somewhat lower power consumption
and no loss in data plane performance relative to analog or hybrid architectures.


The rest of the paper is organized as follows. In Section \ref{sec2} we look at the power consumption of the various beamforming architectures and compare it with the low resolution front end. In \ref{sec3} we present the system model under consideration. We analytically study the effect of low quantization on the reception in Section \ref{sec4}. Simulation results along with discussions are presented in Section \ref{sec5}. Section \ref{sec6} concludes the paper.

\section{Power consumption for digital front ends} \label{sec2}

Before studying the effect of low-resolution 
quantization on capacity, we need
to determine what quantization levels can be implemented in fully digital
architectures in terms of power consumption.
The ADC power consumption scales exponentially by the number of the
quantization bits as \cite{Walden:99}
\[
P_{\rm ADC}  = c f_{s} 2^{n}
\]
where $f_s$ is the sampling rate and $n$ the number of
bits used for quantization, i.e., the ADC resolution and $c$ is the figure of merit (FoM) of the ADC in Joules per conversion step.  
Therefore, by using a small $n$ we can reduce the power consumption of the ADC.

In a recent work \cite{nasri2016700uw}, it is shown that low resolution (4 bit) ADC with FoM of 65 fJ/conversion step can be designed using 65 nm CMOS technology. Leveraging this work and using the power consumption results reported in \cite{Yu201060GHz,kraemer2011design} for low noise amplifiers (LNAs), phase shifters (PS), combiners and mixers we present a comparison of the power consumed by various Rx front ends with 16 antennas in Table \ref{tab:power}.

We see that at 4 bits per ADC, the ADC power consumption is negligible
(33~mW) in comparison to the other RX components such as the LNAs.  Moreover,
at these quantization levels, the lower resolution fully digital BF solution is
actually lower in power consumption than analog or hybrid beamforming architectures,
since the analog and hybrid BF systems require phase shifters and active combiners
not required in the fully digital solution. For the remainder of the study,
we will assume that the fully digital solution can afford at least 4 bits per ADC.



\begin{table} [!t]
\centering
\begin{tabular}{||p{0.7cm}| p{0.6cm}| p{0.6cm}| p{0.6cm}| p{0.6cm}| p{0.7cm}| p{0.7cm}| p{0.7cm}||} \hline\hline
BF & LNA & PS & Comb. & Mixer & ADC (8bits) & ADC (4bits)& Total  \\\hline
Analog & 624 & 312 & 19.5 & 16.8 & 33.3 & -- & 1005.8 \\ \hline
Hybrid (K=2) & 624 & 624 & 39 & 33.2 & 66.6 & -- & 1386.8 \\ \hline
Digital (High res.) & 624 & -- & -- & 268.8 & 532.5 & -- & 1425.8 \\ \hline
Digital (Low res) & 624 & -- & -- & 268.8 & -- & 33.3 & 926.1 \\ \hline\hline
\end{tabular}
\caption{Power consumption (in mW) for each component in the RF chain for various receiver architectures (with 16 Rx antennas).}
\label{tab:power}
\end{table}

\section{System Model} \label{sec3}

We consider a downlink (DL) multiuser small-cell mmWave
communication system 
with ${K}$ base stations (BS). 
Each BS serves up to ${M}$ users
over a total bandwidth of $W_{\rm tot}$.
Each BS is equipped with an antenna array of size $N_{\rm BS}$.
On the other hand, each UE has an array with $N_{\rm UE}$ antenna elements. 

We assume that both the transmitters and the receivers employ
a fully digital architecture with low resolution analog-to-digital (ADC)
and digital-to-analog converters (DAC) as shown in Fig.\ref{fig:txrx}.
To simplify our analysis, we assume a single stream
communication link, i.e., the BSs and the UEs use their arrays to beamform towards each other and boost the effective SINR rather than establish parallel streams.
One may extend our analysis to multi-stream links through eigen-decomposition of the channel and eigen-beamforming.

At each UE, we assume that 
the DL received signal vector $\ybf \in \C^{N_{\rm UE}}$ 
on each sample is given as
\beq\label{eq:noQsig}
\ybf  = \Hbf \wbf_{\rm BS} x + \zbf + \nbf, 
\eeq
where $x \in \mathbb{C}$ is the transmitted symbol,
$\Hbf \in \C^{N_{\rm UE} \times N_{\rm BS}}$
is the (flat fading) 
spatial channel from the transmitter (BS) to the receiver (UE),
and $\wbf_{\rm BS}$ is the transmitter beamforming vector.
The channel matrix $\Hbf$ may vary with time.
The effect of thermal noise and hardware impairments are represented by
the additive Gaussian zero mean vector $\nbf \in \mathbb{C}^{N_{UE}}$.
The elements of $\nbf$ are assumed to be zero mean i.i.d with covariance matrix $\sigma_n^2  \Ibf_{N_{\rm UE}}$.

Due to the high isotropic path loss in mmWave,
cell radius will be typically in the order of a few hundred meters.
Thus it is possible that users associated with different BSs
are in close proximity to each other.
In this case, two or more BSs may beamform towards the same
direction and create a strong interference at a particular set of UEs.
In \ref{eq:noQsig}, the vector $\zbf \in \C^{N_{\rm UE}}$
represents this interfering signal at a given UE.
We assume this interference to be
i.i.d Gaussian with covariance $\sigma_z^2 \Ibf_{N_{\rm UE}}$. 

\section{Link-Layer AQNM Model} \label{sec4}
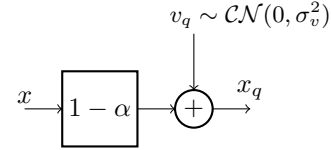
\begin{figure}
\centering
\begin{tikzpicture}[scale=1]
\draw[->] (4,1) -- (4.5,1) node [align=center] at (4,1.15) {$x$};
\draw[thick] (4.5,0.5) rectangle (5.5,1.5) node [align=center] at (5,1) {$1-\alpha$};
\draw[->] (5.5,1) -- (6,1);
\draw[thick] (6.25,1) circle (0.25) node [align=center] at (6.25,1) {$+$};
\draw[->] (6.25,2) -- (6.25,1.25) node [align=center] at (7,2.25) {\small $v_q \sim \mathcal{CN}(0,\sigma_v^2)$};
\draw[->] (6.5,1) -- (7,1) node [align=center] at (7,1.25) {$x_q$} ;
\end{tikzpicture}
\caption{Additive quantization noise (AQN) model for low resolution quantizer.}
\label{fig:aqnm}
\end{figure}

\subsection{Effective SINR}
We first derive a simple analytic model for the effective SINR
as a result of quantization in a multi-antenna receiver.
For this purpose, we use a slightly modified version of the additive
quantization noise model (AQNM) as presented in \cite{FletcherRGR:07}.
In this model in, 
the effect of finite uniform quantization of a scalar input $y$,
is represented as a constant gain plus an additive white
Gaussian noise, as shown in Fig. \ref{fig:aqnm}.
Specifically, it is shown in \cite{FletcherRGR:07} that
if an input complex sample $y$ is modeled as a random variable,
then the quantizer output $y_q$ can be written as,
\beq \label{eq:Qmodel}
	y_q = Q(y) = (1-\alpha)y + v_q, \quad \Exp|v_q|^2 = \alpha(1-\alpha)\Exp|y|^2,
\eeq
where $Q(\cdot)$ denotes the quantization operation and
$v_q$ represents quantization errors uncorrelated with $y$.
The parameter $\alpha \in [0,1]$ is the \emph{inverse coding gain} of the quantizer and is assumed to depend only on the resolution of the quantizer and independent of the input distribution.  
When quantizer resolution is infinite $\alpha=0$. In the AQN model,
we approximate $v_q$ as complex Gaussian. 


We can easily extend this model to a multi-antenna receiver.
In our system model shown in Fig. \ref{fig:txrx}, 
each component $y_i$ of the received signal $\ybf$ is independently
quantized by an ADC before an appropriate receiver-side beamforming
vector $\wbf_{\rm UE}$ is applied.
Thus, from \eqref{eq:noQsig} and \eqref{eq:Qmodel},
the quantized received vector is given as
\beq \label{eq:Qsig}
\ybf_Q  = \Qbf (\ybf) = (1 - \alpha) \Hbf \wbf_{\rm BS} x + (1 - \alpha) \zbf + (1 - \alpha)\nbf + \vbf.
\eeq
The vector $\vbf$ denotes the additive quantization noise (AQN) with covariance 
$\sigma_v^2 \Ibf_{N_{\rm UE}}$.  We assume that the quantization errors
across antennas are uncorrelated.  Using
\eqref{eq:noQsig}, the average per component energy to the
input $\ybf$ of the quantizer is
\begin{align}
	\frac{1}{N_{\rm UE}}\Exp\|\ybf\|^2 &= 
    E_s +  \sigma_{\zbf}^2 + \sigma_n^2,  
\end{align}
where $E_s = (1/N_{\rm UE})\Exp\|\Hbf \wbf_{BS}x\|^2$ is the average
received symbol energy per antenna. 
From \eqref{eq:Qmodel}, the quantization noise
variance is 
\beq \label{eq:sigvq}
	\sigma_v^2 = \alpha(1-\alpha)\left[ E_s + \sigma_{z}^2 + \sigma_n^2 \right].
\eeq
Applying a receiver-side beamforming $\wbf_{\rm UE}$, the channel between the UE and the BS becomes an effective SISO channel.  Define the Rx side BF gain as
\[
	G_{\rm UE} := \Exp|\wbf_{\rm UE}^*\Hbf\wbf_{\rm BS} x|^2 / E_s,
\]
which is the ratio of the signal energy after beamforming to the received signal 
energy per antenna.  
Observe that, if there was no quantization error (i.e.\ $\alpha=0$), the 
post-beamforming SINR would be
\beq \label{eq:gamBF}
	\gamma^{\rm BF} := \frac{\Exp|\wbf_{\rm UE}^*\Hbf\wbf_{\rm BS} x|^2}{
    	\sigma^2_n + \sigma^2_z} = \frac{G_{\rm UE}E_s}{\sigma^2_n + \sigma^2_z}.
\eeq
Now, with quantization, the signal after beamforming is given by
\begin{align}\label{eq:QsigBF}
	\MoveEqLeft y_Q^{\rm BF} := \wbf_{\rm UE}^* \ybf_Q 
    = (1 - \alpha) \wbf_{\rm UE}^* \Hbf \wbf_{\rm BS} x \nonumber \\
	& + (1 - \alpha) \wbf_{\rm UE}^* \zbf + (1 - \alpha) \wbf_{\rm UE}^* \nbf + 
    \wbf_{\rm UE}^* \vbf. 
\end{align} 
Without loss of generality, we assume
$\|\wbf_{\rm UE}\| = 1$.  Then, the average received signal energy post-beamforming is
\beq \label{eq:Esq}
	E_s^{\rm BF} = (1-\alpha)^2\Exp|\wbf_{\rm UE}^*\Hbf\wbf_{\rm BS}x|^2 = 
    G_{\rm UE}E_s,
\eeq
while the average noise energy is
\beq \label{eq:Enq}
	E_{\rm n} = (1-\alpha)^2(\sigma^2_z + \sigma^2_n) + \sigma^2_v.
\eeq
Combining \eqref{eq:sigvq}, \eqref{eq:gamBF},
\eqref{eq:Esq} and \eqref{eq:Enq}, we obtain that
the SINR after beamforming is
\beq \label{eq:gamQBF}
	\gamma^{\rm BF}_Q = \frac{E_s^{\rm BF}}{E_{\rm n}} =
    \frac{(1-\alpha)\gamma^{\rm BF}}{1 + (\alpha/G_{\rm UE}) \gamma^{\rm BF}}.
\eeq

\subsection{Regimes}
Using  \eqref{eq:gamQBF}, we can immediately qualitatively
understand the system-level effects
of quantization by looking at two regimes:  
In the low-SINR regime ($\gamma^{\rm BF}$ is small),
\beq
	\gamma^{\rm BF}_Q \approx (1-\alpha)\gamma^{\rm BF},
\eeq
so the SINR is decreased only by a factor $1-\alpha$.  We will see that at moderate
quantization levels, this decrease is extremely small.
In the high SINR regime $\gamma^{\rm BF} \arr \infty$,
\beq
	\gamma^{\rm BF}_Q \arr \frac{G_{\rm UE}(1-\alpha)}{\alpha}.
\eeq
Thus, the effect of quantization is to saturate the maximum SINR.
Hence, we conclude that the effect of using low-resolution quantization
is essentially on the maximum SINR.
We will see that, at practical quantization levels
and beamforming gains, this SINR limit is not significant in cellular systems, which are more limited by low SINR mobiles than high SINR ones.

\subsection{Validating AQNM}

\begin{table}
\begin{center}
\begin{tabular}{|>{\raggedright}p{3 cm}|>{\raggedright}p{4 cm}|}
  \hline
  {\bf Parameter} & {\bf Value}
  \tabularnewline \hline

Used bandwidth, $W$ & 90 MHz
\tabularnewline \hline

FFT size, $N_{\rm fft}$ & $2048$
\tabularnewline \hline

Subcarrier spacing  & $75$~kHz
\tabularnewline \hline

Used subcarriers, $N_{\rm sc}$ & $1200$
\tabularnewline \hline

Symbol duration & Type 0 $14.3750$ , Type 1 $14.2708$
\tabularnewline \hline

Slot duration & $100$~${\rm \mu}$s
\tabularnewline \hline

\end{tabular}
\caption{AQNM validation OFDM parameters}
\label{tab:aqnSimPar}
\end{center}
\end{table}

\begin{figure}[t!]
\centering
\includegraphics[width=0.45\textwidth]{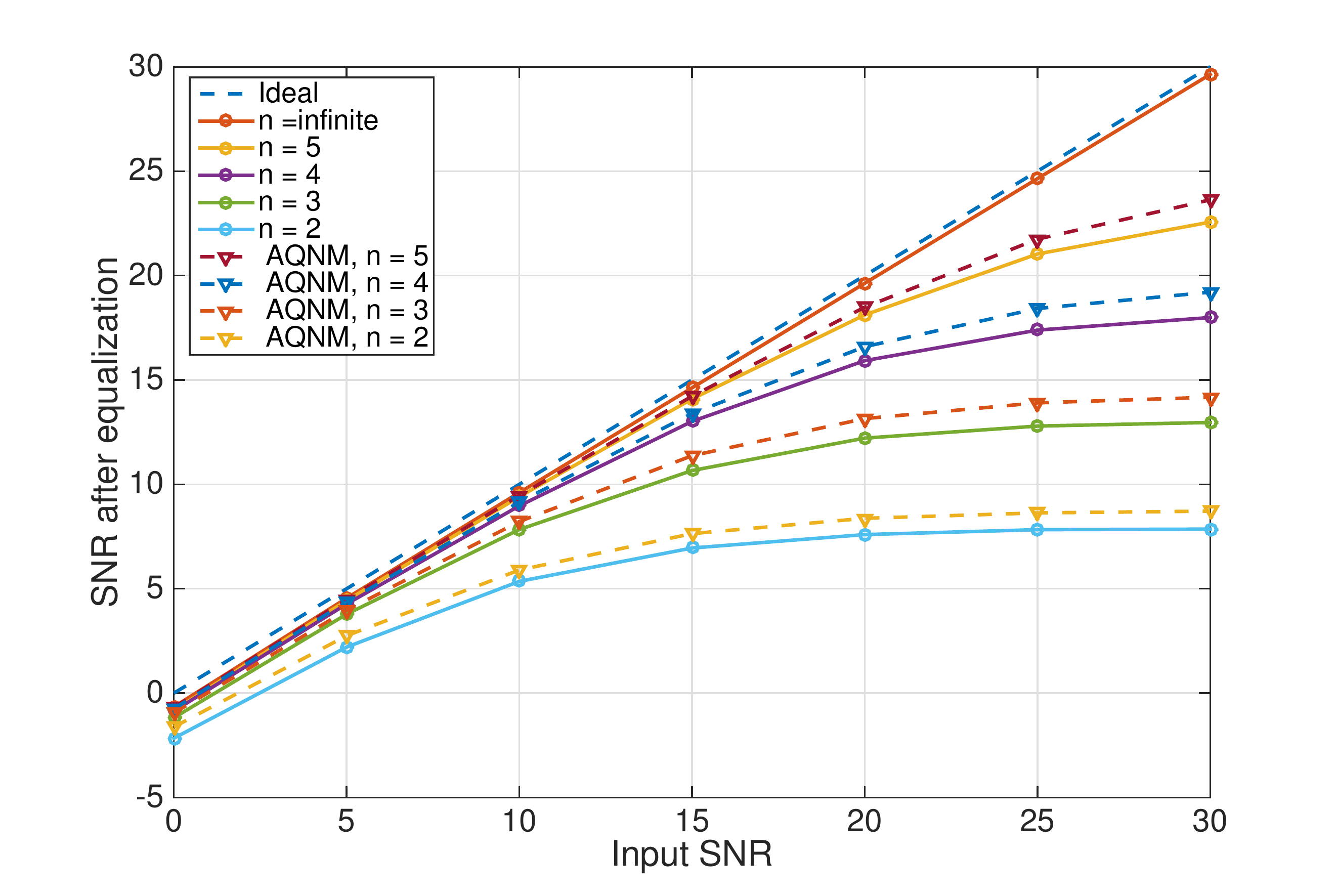}
    \caption{ Verifying the accuracy of the AQN model.
    The model follows simulated quantization (solid lines) well.
    At ${\rm 20}$~dB SNR and for all resolution bits,
    the distance between simulated
    and linear model quantization is below ${\rm 1}$~dB. }
    \label{fig:linearqSNR}
\end{figure}

\begin{figure}[t!]
\centering
\includegraphics[width=0.45\textwidth]{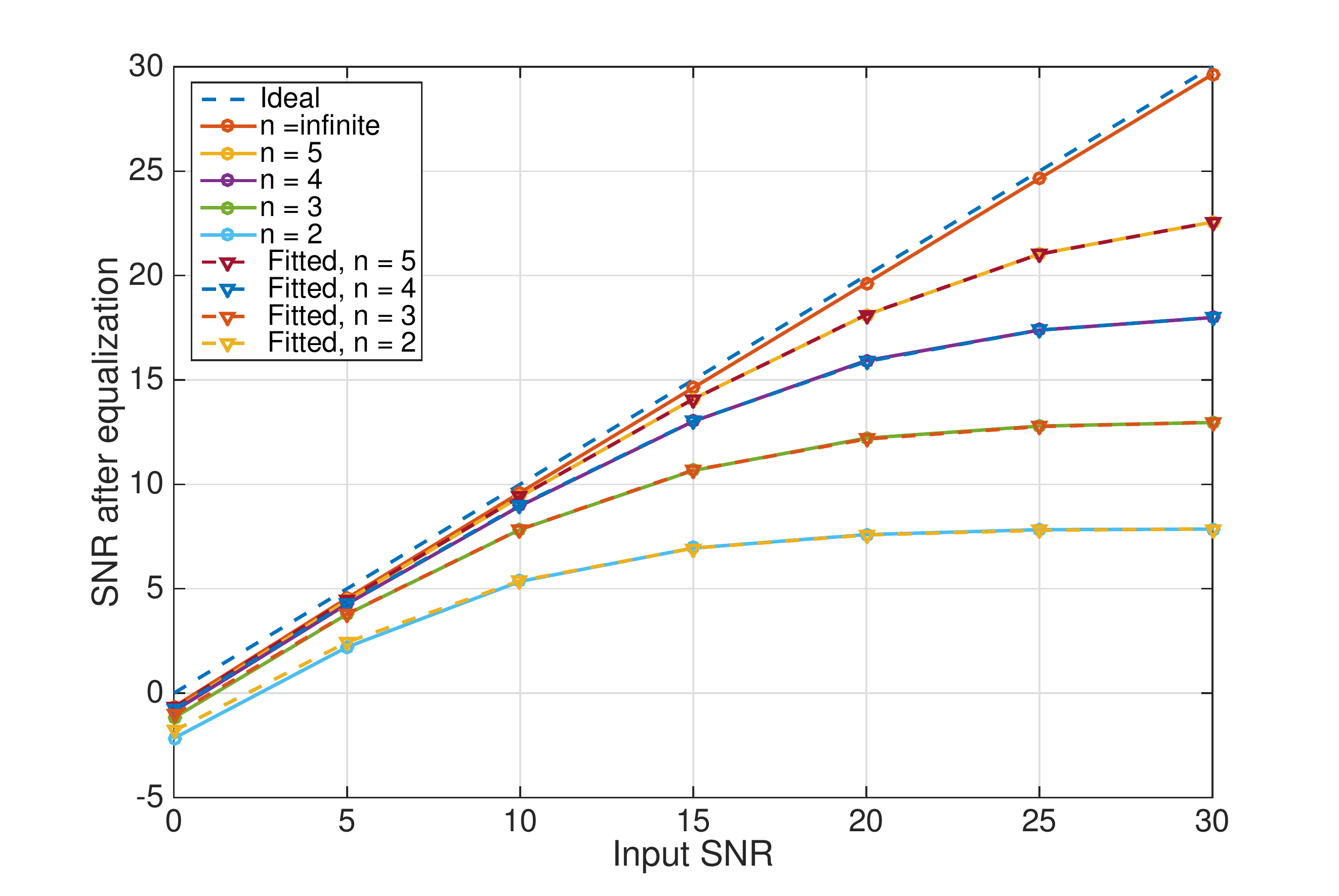}
    \caption{Quantization model fitted to data. 
    The model follows
    simulated quantization (solid lines) perfectly.}
    \label{fig:linqSNRfit}
\end{figure}

In order to validate the accuracy of the effective SINR model \eqref{eq:gamQBF}, 
we compare the effective output SNR predicted by the AQN model 
\eqref{eq:gamQBF} with the actual post-equalization SNR obtained by a
detailed single link OFDM system.
In our link level OFDM simulator, the transmitter generates random complex
symbols and modulates them into OFDM symbols using the parameters in Table \ref{tab:aqnSimPar}
from the 5G Verizon specification \cite{verizon5g2016}. 
We assume an AWGN channel where at the receiver the symbols are filtered, quantized, equalized and demodulated using known reference signals for channel estimation.

Fig. \ref{fig:linearqSNR} compares the effective SNR
predicted by the AQN model \eqref{eq:gamQBF}
with the simulated post-equalization SNR, for varying quantizer resolutions ($n$). The value of $\alpha$ is computed assuming an optimal uniform $n$-bit quantizer. We observe that the theoretical model predicts a close approximation of the post-equalization SNR. 
Importantly, we see that the quantization has the effect of saturating the SINR.
In particular, if we look at $n \geq 3$, 
it is clear that at SNRs below $15$ dB, the effect of quantization noise on the system is negligible.

Nevertheless, the model \eqref{eq:gamQBF} slightly over-predicts the effective SINR
since the actual SINR is degraded by other factors including channel estimation 
error.  To account for these losses,
in Fig. \ref{fig:linqSNRfit} we plot the prediction of the AQN model when $\alpha$ obtained through a least squares fit to the simulated output SNR. This shows that the slight over-estimation seen in Fig. \ref{fig:linearqSNR} can be avoided by using a better optimization algorithm for the computation $\alpha$.

\begin{figure}
\centering
 \includegraphics[scale=0.22]{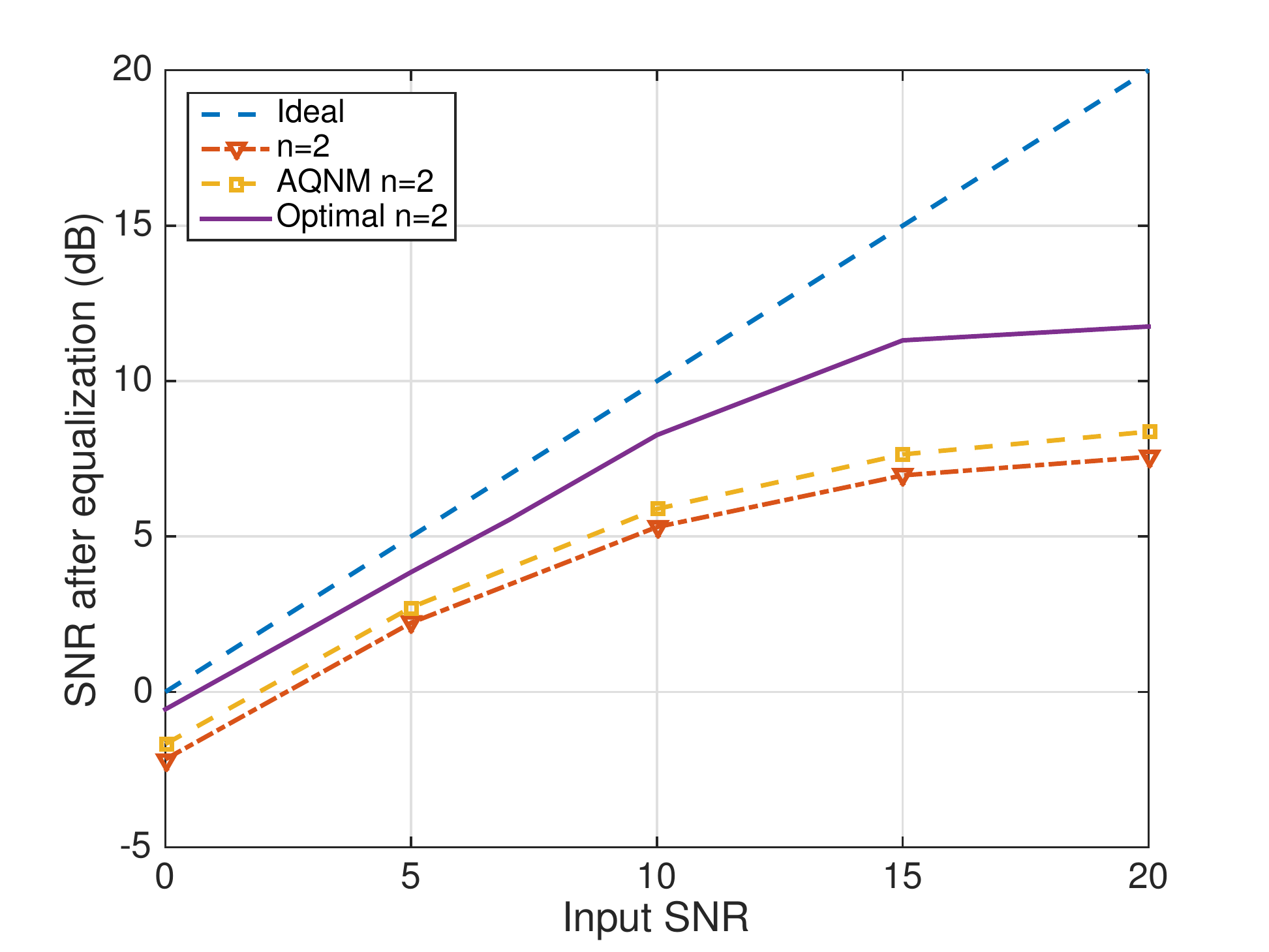}
 \includegraphics[scale=0.22]{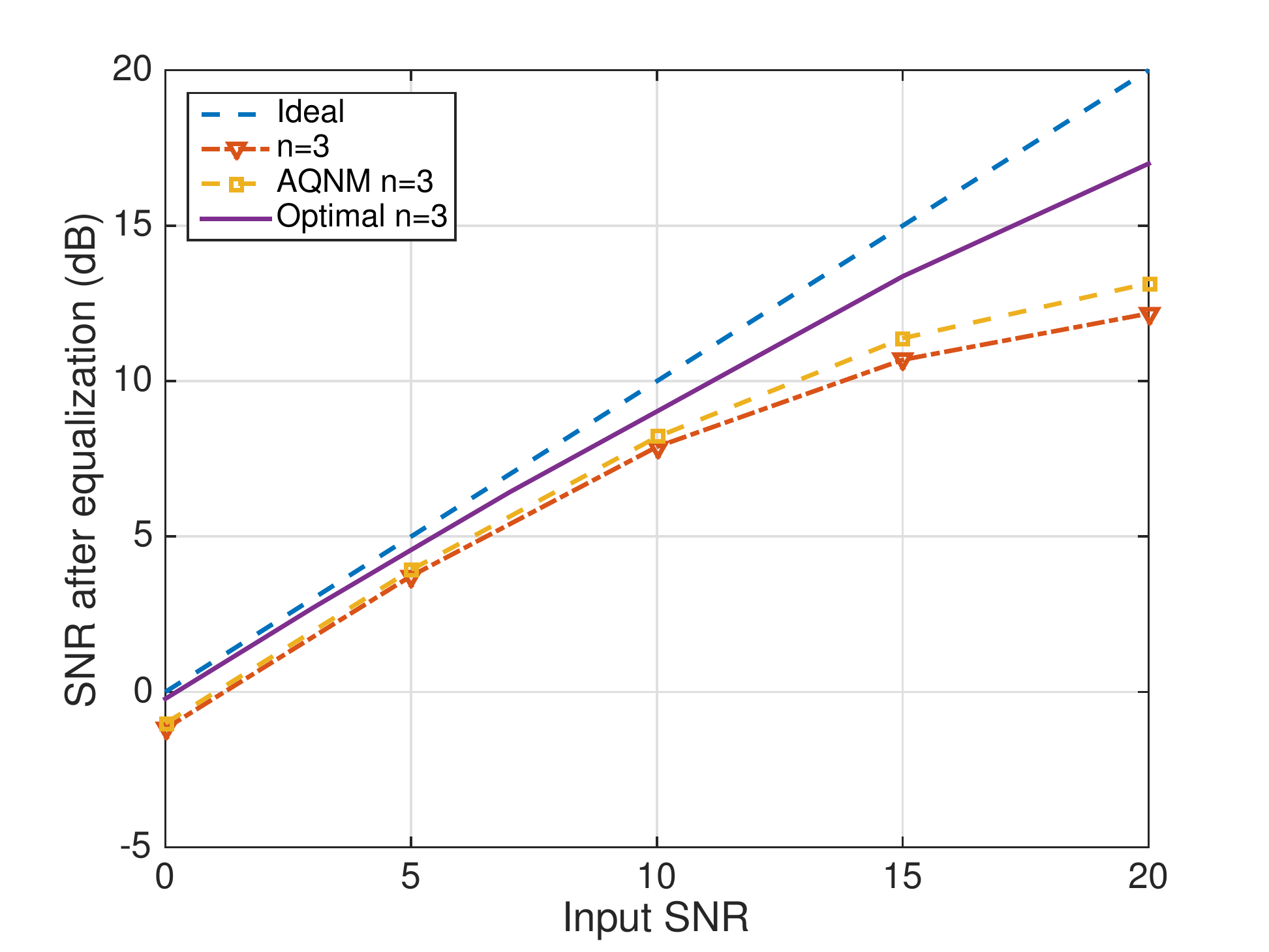}
 \caption{Output SNR predicted by the AQNM model compared with the achievable SNR with optimal signal constellation.}
 \label{fig_opt}
\end{figure}

Note that all the above calculations have assumed linear Gaussian quantization 
noise, which is equivalent to performing linear processing.  Theoretically,
it is possible to improve the capacity with nonlinear processing.
To quantify the maximum possible gain,
Fig. \ref{fig_opt},
compares the post equalization SNR with the information theoretic 
SNR achieved with an optimum input signal constellation as derived in 
\cite{Singh2009Limits}.
We observe that in the low SNR regime the deviation from the optimum is less than $1$~dB for $n=2$ and less than $0.5$~dB for $n=3$. Although, this gap scales with the operating SNR.
We conclude that, at low SNRs, linear processing does not significantly reduce the link-layer capacity.

\section{Downlink System Capacity}\label{sec5}


\begin{figure}[!t]
  \centering  
    \includegraphics[width=0.45\textwidth]{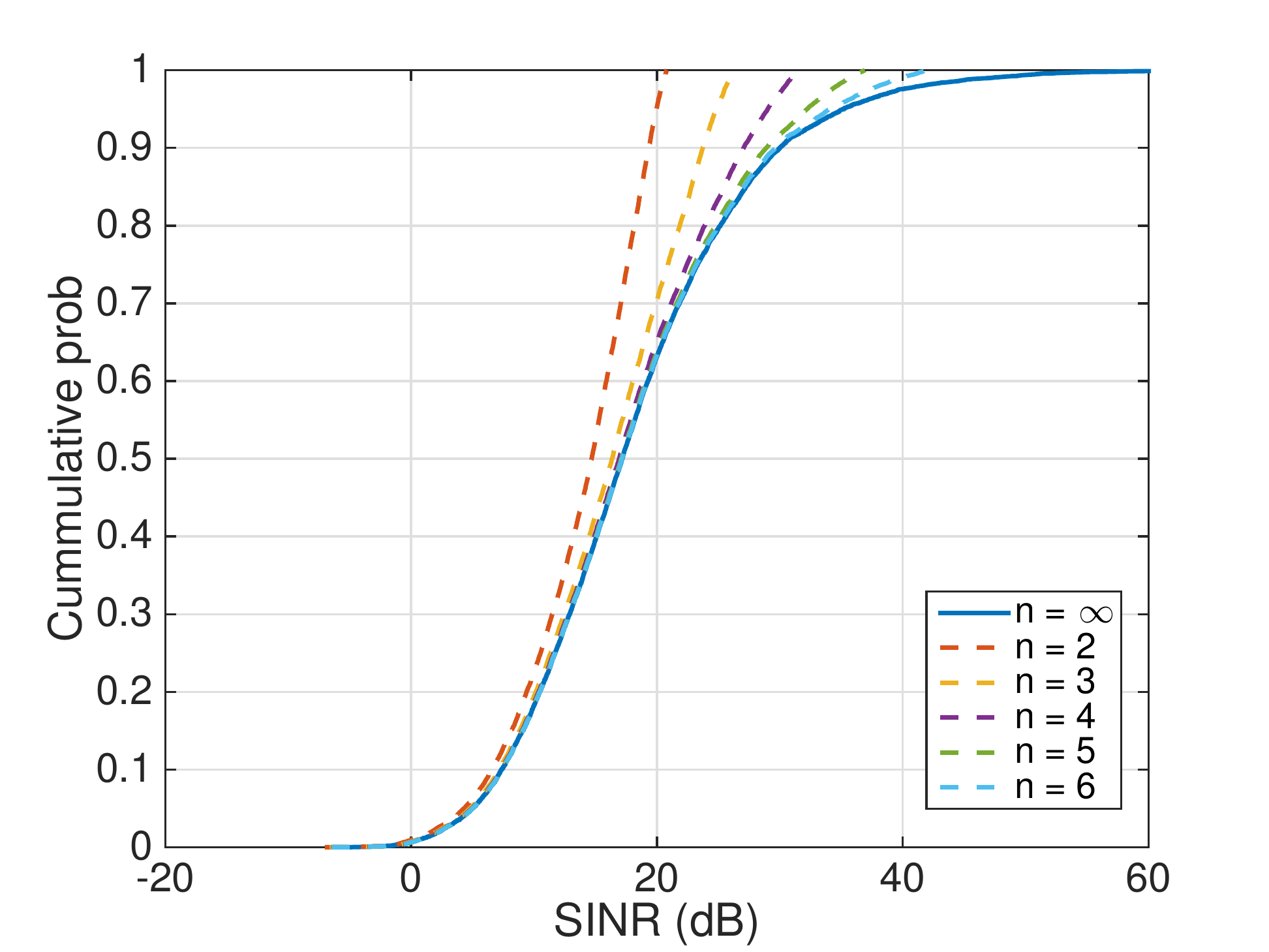}
    \caption{SINR distribution for various quantization resolutions compared with the distribution assuming infinite resolution quantizer.}
    \label{fig:sysSinr}
\end{figure}

\begin{table}
\begin{center}
\begin{tabular}{|>{\raggedright}p{3 cm}|>{\raggedright}p{4 cm}|}
  \hline
  {\bf Parameter} & {\bf Description}
  \tabularnewline \hline



Cell radius & $100$~m
\tabularnewline \hline

Pathloss model & \cite{AkdenizCapacity:14}
\tabularnewline \hline

Carrier frequency & $28$~GHz
\tabularnewline \hline

DL bandwidth $(W_{\rm tot})$ & $1$~GHz
\tabularnewline \hline

DL Tx power & $30$~dBm
\tabularnewline \hline

Rx noise figure & $7$~dB
\tabularnewline \hline

BS antenna array & $8 \times 8$ uniform planar
\tabularnewline \hline

UE antenna array & $4 \times 4$ uniform planar
\tabularnewline \hline

BF mode & Digital long-term, single stream
\tabularnewline \hline

\end{tabular}
\caption{Multicell simulation parameters}
\label{tab:cellSimPar}
\end{center}
\end{table}

We conclude by applying our link-layer AQN model \eqref{eq:gamQBF},
to understand the effect of low-resolution quantization in the downlink 
system capacity \cite{3GPP36.814}.
We simulate a $2$km by $2$km area covered by a multiplicity of hexagonal cells with each cell being divided into three sectors.
Each sector is assumed to serve on average $10$~UEs which
are randomly ``dropped''.
We then compute a random path loss between the BS and the UEs based on the urban channel model presented in \cite{AkdenizCapacity:14}.
We simulate a DL transmission scenario where BSs transmit a single stream to every user.
Both BSs and UEs are assumed to perform longterm digital beamforming
\cite{Lozano:07}
making use of the spatial second-order statistics of channel.
The parameters of this simulation are summarized in Table \ref{tab:cellSimPar}.


The random placement of the users makes it is possible for two or more BS-UE pairs to beamform in the same direction causing inter-cell interference.
Under the assumption of AQN, we use
\eqref{eq:gamQBF} to compute the post beamforming SINRs for a given
value of ADC resolution $n$.
To demonstrate the effect of low resolution we vary $n$ from 2 -- 6 and
plot the distribution of the SINR in
Fig.~\ref{fig:sysSinr} along with the curve for infinite resolution
($n=\infty$).

From Fig. \ref{fig:sysSinr} we observe that at low SINRs, the deviation
from the infinite resolution curve is minimal if any.
On the other hand, at high SINR regimes we observe a ``clipping'' of the maximum achievable SINR. 
More specifically, the SINR penalty for 2 bit quantization is less than
$10$~dB for the $90$-th percentile and above, while the same loss is
$6$~dB for 3 bit resolution. In the $50$-th percentile on the other
hand this difference drops at about $2.5$ and $1$~dB respectively. 

Obtaining the effective SINR using the theoretical model, we next plot
the theoretically achievable rates under various quantizer resolutions.
Following the analysis in \cite{AkdenizCapacity:14} and the
link-layer model \cite{MogEtAl:07}, we assume a $3~$dB loss from Shannon capacity,
a 20\% overhead and a maximum spectral efficiency of $\rho=5.5$ bps/Hz
The loss due to quantization is not
noticeable for $n \geq 3$
The reason is twofold.
Firstly, very few
users in the system will operate at high SINR, thus the clipping of SINR
as observed in Fig. \ref{fig:sysSinr} has a small effect on the average
rate.
Secondly, as rate is a logarithmic function of the SINR,
increasing the SINR beyond a certain point produces diminishing
increase in the rate, particularly with the maximum spectral efficiency.

\begin{figure}[!t]
\centering
    \includegraphics[width=0.45\textwidth]{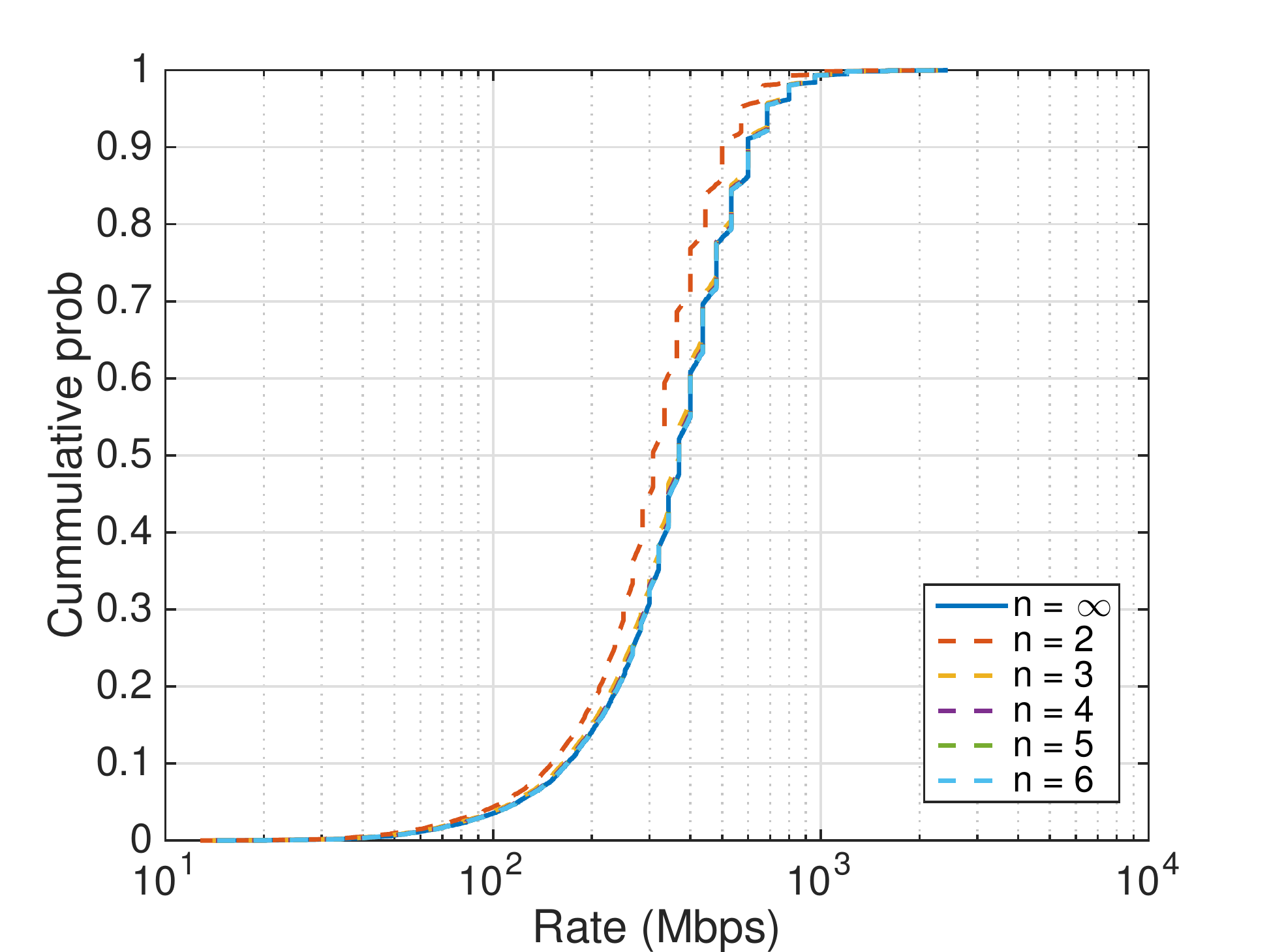}
    \caption{ Rate distribution for various quantization resolutions compared with the distribution assuming infinite resolution quantizer.}
    \label{fig:sysCap}
\end{figure}

\section{Conclusions} \label{sec6}
We have presented a simple analytic method for evaluating 
low resolution quantization in cellular system capacity based on an AQN model.
Qualitatively, the AQN model shows that the finite quantization has the
effect of saturating the maximum SINR.
Using this model, we show that
mmWave systems that use high gain beamforming are unlikely to be significantly 
impaired by these limits when the quantization levels are set to 3--4 bits.
These quantization resolutions are entirely supportable with current state-of-the-art ADCs.
These findings indicate that fully digital BF architectures 
can provide comparable data plane performance as analog BF,
with similar or lower power consumption, while also obtaining all the benefits in the control plane.



%

\bibliographystyle{IEEEtran}
\bibliography{bibl}

\newcommand{\SortNoop}[1]{}
\begin{thebibliography}{10}
\providecommand{\url}[1]{#1}
\csname url@samestyle\endcsname
\providecommand{\newblock}{\relax}
\providecommand{\bibinfo}[2]{#2}
\providecommand{\BIBentrySTDinterwordspacing}{\spaceskip=0pt\relax}
\providecommand{\BIBentryALTinterwordstretchfactor}{4}
\providecommand{\BIBentryALTinterwordspacing}{\spaceskip=\fontdimen2\font plus
\BIBentryALTinterwordstretchfactor\fontdimen3\font minus
  \fontdimen4\font\relax}
\providecommand{\BIBforeignlanguage}[2]{{%
\expandafter\ifx\csname l@#1\endcsname\relax
\typeout{** WARNING: IEEEtran.bst: No hyphenation pattern has been}%
\typeout{** loaded for the language `#1'. Using the pattern for}%
\typeout{** the default language instead.}%
\else
\language=\csname l@#1\endcsname
\fi
#2}}
\providecommand{\BIBdecl}{\relax}
\BIBdecl

\bibitem{RanRapE:14}
S.~Rangan, T.~S. Rappaport, and E.~Erkip, ``Millimeter-wave cellular wireless
  networks: Potentials and challenges,'' \emph{Proc. IEEE}, vol. 102, no.~3,
  pp. 366--385, Mar. 2014.

\bibitem{KhanPi:11-CommMag}
F.~Khan and Z.~Pi, ``{An introduction to millimeter-wave mobile broadband
  systems},'' \emph{IEEE Commun. Mag.}, vol.~49, no.~6, pp. 101 -- 107, Jun.
  2011.

\bibitem{Zhang:05}
X.~Zhang, A.~Molisch, and S.-Y. Kung, ``Variable-phase-shift-based
  {RF}-baseband codesign for {MIMO} antenna selection,'' \emph{IEEE Trans.
  Signal Process.}, vol.~53, no.~11, pp. 4091--4103, Nov. 2005.

\bibitem{barati2015directional}
C.~Barati~Nt., S.~Hosseini, S.~Rangan, P.~Liu, T.~Korakis, S.~Panwar, and T.~S.
  Rappaport, ``Directional cell discovery in millimeter wave cellular
  networks,'' \emph{IEEE Trans. Wireless Commun.}, vol.~14, no.~12, pp. 6664 --
  6678, Nov. 2015.

\bibitem{barati2016initial}
C.~N. Barati, S.~A. Hosseini, M.~Mezzavilla, T.~Korakis, S.~S. Panwar,
  S.~Rangan, and M.~Zorzi, ``Initial access in millimeter wave cellular
  systems,'' \emph{IEEE Trans.\ Wireless Commun}, vol.~15, no.~12, pp.
  7926--7940, Dec 2016.

\bibitem{dutta2017frame}
S.~Dutta, M.~Mezzavilla, R.~Ford, M.~Zhang, S.~Rangan, and M.~Zorzi, ``Frame
  structure design and analysis for millimeter wave cellular systems,''
  \emph{IEEE Trans.\ Wireless Commun}, vol.~16, no.~3, pp. 1508--1522, Mar.
  2017.

\bibitem{Singh2009Limits}
J.~Singh, O.~Dabeer, and U.~Madhow, ``On the limits of communication with
  low-precision analog-to-digital conversion at the receiver,'' \emph{IEEE
  Trans. Commun.}, vol.~57, no.~12, pp. 3629--3639, Dec. 2009.

\bibitem{oner2015adc}
O.~Orhan, E.~Erkip, and S.~Rangan, ``Low power analog-to-digital conversion in
  millimeter wave systems: Impact of resolution and bandwidth on performance,''
  in \emph{Proc. Information Theory and Applications Workshop (ITA)}, Feb.
  2015, pp. 191--198.

\bibitem{mo2014channel}
J.~Mo, P.~Schniter, N.~G. Prelcic, and R.~W. Heath~Jr, ``Channel estimation in
  millimeter wave {MIMO} systems with one-bit quantization,'' in \emph{Proc.\
  Asilomar Conf. on Signals, Systems and Computers}, Nov. 2014.

\bibitem{mo2014high}
J.~Mo and R.~W. Heath, ``High {SNR} capacity of millimeter wave {MIMO} systems
  with one-bit quantization,'' in \emph{Proc.\ Information Theory and
  Applications Workshop (ITA)}, Feb. 2014, pp. 1--5.

\bibitem{Walden:99}
R.~H. Walden, ``Analog-to-digital converter survey and analysis,'' \emph{IEEE
  J. Sel. Areas Commun.}, vol.~17, no.~4, pp. 539--550, Apr. 1999.

\bibitem{nasri2016700uw}
B.~Nasri, S.~P. Sebastian, K.~D. You, R.~RanjithKumar, and D.~Shahrjerdi, ``A
  700 {uW} 1{GS/s} 4-bit folding-flash {ADC} in 65nm {CMOS} for wideband
  wireless communications,'' in \emph{Proc. ISCAS}, May 2017, pp. 1--4.

\bibitem{Yu201060GHz}
Y.~Yu, P.~G.~M. Baltus, A.~de~Graauw, E.~van~der Heijden, C.~S. Vaucher, and
  A.~H.~M. van Roermund, ``A 60 ghz phase shifter integrated with lna and pa in
  65 nm cmos for phased array systems,'' \emph{IEEE J. Solid-State Circuits},
  vol.~45, no.~9, pp. 1697--1709, Sept 2010.

\bibitem{kraemer2011design}
M.~Kraemer, D.~Dragomirescu, and R.~Plana, ``Design of a very low-power,
  low-cost 60 ghz receiver front-end implemented in 65 nm cmos technology,''
  \emph{Int. J. of microwave and wireless technologies}, vol.~3, no.~2, pp.
  131--138, 2011.

\bibitem{FletcherRGR:07}
A.~K. Fletcher, S.~Rangan, V.~K. Goyal, and K.~Ramchandran, ``Robust predictive
  quantization: Analysis and design via convex optimization,'' \emph{IEEE J.
  Sel. Topics Signal Process.}, vol.~1, no.~4, pp. 618--632, Dec. 2007.

\bibitem{verizon5g2016}
{TS V5G.211}, ``Verizon 5th generation radio access {(V5G RA)}; physical
  channels and modulation,'' Rel. 1, 2016, available on-line at {\tt
  http://www.5gtf.net}.

\bibitem{AkdenizCapacity:14}
M.~Akdeniz, Y.~Liu, M.~Samimi, S.~Sun, S.~Rangan, T.~Rappaport, and E.~Erkip,
  ``Millimeter wave channel modeling and cellular capacity evaluation,''
  \emph{IEEE J. Sel. Areas Commun.}, vol.~32, no.~6, pp. 1164--1179, June 2014.

\bibitem{3GPP36.814}
3GPP, ``{Further advancements for {E-UTRA} physical layer aspects},'' TR 36.814
  (release 9), 2010.

\bibitem{Lozano:07}
A.~Lozano, ``Long-term transmit beamforming for wireless multicasting,'' in
  \emph{Proc.\ ICASSP}, vol.~3, Apr. 2007, pp. III--417--III--420.

\bibitem{MogEtAl:07}
P.~Mogensen, W.~Na, I.~Z. Kov{\'a}cs, F.~Frederiksen, A.~Pokhariyal, K.~I.
  Pedersen, T.~Kolding, K.~Hugl, and M.~Kuusela, ``{LTE} capacity compared to
  the {S}hannon bound,'' in \emph{Proc. IEEE 65th Vehicular Technology
  Conference (VTC)}, April 2007, pp. 1234--1238.

\end{thebibliography}

\vspace{-1em}



\end{document}